\documentclass[12pt]{article}
\usepackage{epsfig}
\usepackage{latexsym}
   \title{Quantum Corrections to the Reissner-Nordstr\"{o}m and Kerr-Newman 
   Metrics}
   \author{John F. Donoghue$^a$ and Barry R. Holstein$^{a,b}$\\
   $^a$Department of Physics\\
   University of Massachusetts\\
   Amherst, MA  01003\\
   and\\
   $^b$Institut f\"{u}r Kernphysik\\
   Forschungszentrum J\"{u}lich\\
   D-52425 J\"{u}lich, Germany\\
   and\\
   Bj\"{o}rn Garbrecht and Thomas Konstandin\\
   Fakult\"{a}t f\"{u}r Physik und Astronomie\\
   Universit\"{a}t Heidelberg\\
   D-69120 Heidelberg, Germany}
   
\begin{document}
   \begin{titlepage}
   \maketitle
   \begin{abstract}
   We use effective field theory techniques to examine the quantum corrections 
   to the gravitational 
   metrics of charged particles, with and without spin. In momentum space
   the masslessness of the photon implies
   the presence of nonanalytic pieces $\sim\sqrt{-q^2},q^2\log -q^2$,
   etc. in the form factors of the energy-momentum tensor. We show how the
   former reproduces the classical non-linear terms of the 
Reissner-Nordstr\"{o}m
   and Kerr-Newman
   metrics while the latter can be interpreted as quantum corrections to these
   metrics, of order $G\alpha\hbar/mr^3$.
   \end{abstract}
   \end{titlepage}

   \section{Introduction}
   The gravitational field around a particle is described by a metric, which 
   provides a solution to Einstein's Equation of general relativity. This theory is clearly
a classical field theory. In this paper, we will discuss quantum corrections to
   the
   metric and show that these quantum effects are reliably calculable for
   the case of a charged particle, either with or without spin, using the
techniques of effective field theory\cite{dgh,eft,gl,don}.

   The classical solution of the
   Einstein equations for a massive charged field are described by the
   Reissner-Nordstr\"{o}m metric. In harmonic gauge, this metric has 
   has the form\footnote{Note 
that throughout this paper we use $\alpha=e^2/4\pi$ without any factor
   of 
$\hbar$ so that we may use it to desdribe the classical corrections. 
This does not make a difference in the sections where 
we use relativistic notation with $\hbar=1$, but we adopt this 
convention so that the all factors of $\hbar$ will be visible 
in those equations where we make this constant explicit. We use 
$c=1$ units in all sections.}\cite{rnm}
   \begin{eqnarray}
   g_{00}&=& 1-{2Gm\over r}+{G\alpha\over r^2}+\ldots\nonumber\\
   g_{0i}&=&0\nonumber\\
   g_{ij}&=&-\delta_{ij}-\delta_{ij}{2Gm\over r}+G\alpha{r_ir_j\over
   r^2}+\ldots\label{eq:rn}
   \end{eqnarray}
   The metric for a spinning charge is known as
   the Kerr-Newman metric and its form in harmonic gauge is found to
   be\cite{ker,knm}
   \begin{eqnarray}
   g_{00}&=& 1-{2Gm\over r}+{G\alpha\over r^2}+\ldots\nonumber\\
   g_{0i}&=&({2G\over r^3}-{G\alpha\over
   mr^4})(\vec{S}\times\vec{r})_i\nonumber\\
   g_{ij}&=&-\delta_{ij}-\delta_{ij}{2Gm\over r}+G\alpha{r_ir_j\over
   r^2}+\ldots\label{eq:kn}
   \end{eqnarray}

   We will use effective field theory techniques to recreate both the
   classical terms in these potentials, and also find quantum corrections of the
   following forms.
   For the
   Reissner-Nordstr\"{o}m metric we find 
   \begin{eqnarray}
   g_{00}&=& 1-{2Gm\over r}+{G\alpha\over r^2}
 - {8G\alpha\hbar \over 3\pi mr^3}+\ldots\nonumber\\
   g_{0i}&=&0\nonumber\\
   g_{ij}&=&-\delta_{ij}-\delta_{ij}{2Gm\over r}+G\alpha{r_ir_j\over
   r^2}+{4G\alpha\hbar\over 3\pi mr^3}\left({r_ir_j\over
   r^2}-\delta_{ij}\right)+\ldots
   \end{eqnarray}
   For spin 1/2 particle we reproduce 
   the Kerr-Newman metric with $J = \hbar/2$ plus quantum effects of the form
   \begin{eqnarray}
   g_{00}&=& 1-{2Gm\over r}+{G\alpha\over r^2} 
  - {8G\alpha\hbar \over 3\pi mr^3} +\ldots\nonumber\\
   g_{0i}&=&({2G\over r^3}-{G\alpha\over
   mr^4}+{2G\alpha\hbar\over \pi m^2r^5})(\vec{S}\times\vec{r})_i\nonumber\\
   g_{ij}&=&-\delta_{ij}-\delta_{ij}{2Gm\over r}+G\alpha{r_ir_j\over
   r^2}+ {4G\alpha\hbar\over 3\pi mr^3}\left({r_ir_j\over
   r^2}-\delta_{ij}\right)+\ldots
   \end{eqnarray}
   Notice that the spin-independent terms are universal for both the 
   classical and quantum corrections. 

   We start by discussing the 
general concept of quantum effects in the metric. To the 
   best of our knowledge the content of this discussion 
   has not appeared before in the literature. We review the logic of
   effective field theory that we use to analyse this problem. We then turn to the
   extraction of the specific quantum corrections that are relevant for this
   problem, first 
   for spinless particles and then for particles with spin 1/2.

   \section{Quantum corrections to metrics}

   The use of the metric field in General Relativity is inherently classical. 
   However, there is a well defined context that one can discuss quantum
   corrections
   to the metric. This occurs when it is quantum matter, not quantum gravity, that
   is 
   responsible for the quantum effects, and when matter yields such effects which 
   are larger at large distances than the effects of quantum gravity. We will show
   that 
   the massless nature of the photon implies that these conditions are satisfied
   for the
   metrics around charged particles. In these cases the long
   range 
   propagation of photons provides the quantum modifications of the metric at
   scales where 
   gravity is still classical.  

   We will make use of the logic developed in the study of effective field
   theories. 
   Effective field theory techniques have been developed primarily for situations
   where
   there are multiple scales in the problem and are used to identify effects from
   the physics
   relevant at the lowest energy scales or largest distance scales. In our case,
   the
   lighest scale is the photon mass, i.e. zero energy, so that we will be
   separating
   the effects of massive degrees of freedom from the massless ones\footnote{Note
   that 
   effective field theory techniques are often applied to nonrenormalizable
   theories. 
   That is not our use here as we are dealing with a renormalizable theory, QED.
   However, 
   the logic of effective field theory is still useful in this context.}. 

   Gravity couples to the energy momentum tensor $T_{\mu\nu}$ of a particle. We 
   will calculate the energy momentum tensor in a power series in $\alpha$, using 
   usual Feynman diagram techniques. These calculations are straightforward 
   applications of QED\cite{ber}. 
The result will be expressed in terms of form factors 
   of the various allowed Lorentz structures in the matrix element of $T_{\mu\nu}$.
   Let us here generically call such a form factor $F(q^2)$, with $q_\mu$ being the
   momentem transfer. Specific cases will be presented later. The form factor is
   the 
   momentum space description of the structure of the particle. Because the
   massless
   photon couples to gravity, this form factor has features not common in most
   other
   form factors. Normally, form factors can be expanded in a power series in $q^2$
   around
   $q^2=0$, with the coefficients being related to the structure of the particle.
   For 
   example, the coefficient of the term linear in $q^2$ is related to the 
   ``charge radius squared'' of the particle, i.e.
\begin{equation}
<r^2> = 6{d\over dq^2} F(q^2) |_{q^2=0}
\end{equation}
 These analytic terms in the
   power
   series can equally well be represented by effective Lagrangians with higher
   powers of derivitives of the fields. However, in the gravitational case, 
   photonic diagrams yield {\em non-analytic} terms in the expansion of the form
   factor.
   In particular, we will find square-root and logarithmic non-analytic terms, i.e.
   \begin{equation}
   F(q^2) = 1+ a\alpha 
   {q^2 \over m^2}\sqrt{m^2 \over -q^2} + 
   b\alpha{q^2 \over m^2} \log(-q^2) + c \alpha {q^2 \over m^2} +\ldots
   \end{equation}
   where a,b,c are constants.
   These non-analytic terms cannot be represented by effective Lagrangians
   and can only arise from the long range propagation of massless particles. 
   Note that they imply that the gravitational charge radius is infinite, 
   which reflects the fact that 
   the energy in the electric field extends out to infinity.
   These non-analytic terms generate the effects that we seek\footnote{In 
   chiral
   perturbation theory these are well known as non-analytic dependences on
   the pion mass $\sqrt{m^2_\pi}$ and $\log(m_\pi^2)$\cite{pagli}. The only other physical
   situation
   where there are non-analytic terms in the momentum transfer itself involves
   massless gravitons in the effective field theory of general
   relativity\cite{don}}. 

   The spatial distribution of energy, and hence the metric, will be recovered
   by a Fourier transformation to coordinate space. Generically the position 
   dependent terms in the metric will be 
   \begin{eqnarray}
   {\rm metric} &\sim& Gm\int {d^3q \over (2\pi)^3 } 
 e^{i\vec{q}\cdot\vec{r}} {1 \over \vec{q}^2}
   \left[ 1- b\alpha 
   {\vec{q}^2 \over m^2}\sqrt{m^2 \over \vec{q}^2} - 
   {\vec{q}^2 \over m^2} \log(\vec{q}^2) -c\alpha {\vec{q}^2 \over m^2}+\ldots 
   \right] \nonumber \\
   &\sim& Gm \left[ {1\over r} +{a\alpha \over m r^2} +{ b\alpha \hbar\over
   m^2 r^3}
   +{c\alpha \over m^2} \delta^3(x) +\ldots \right]
   \end{eqnarray}
(Numerical factors of order unity will be inserted later.)
   The leading piece in the form factor yields the usual ``Newtonian''
   component 
   of the metric. The analytic term in the form factor yields a 
   delta function - i.e. no effect at large distances. (Higher order 
   analytic terms produce derivitives of the delta function.) However, the
   two non-analytic terms produce the effects that we are interested in. The 
   square-root generates the classical correction in the metric of order $\alpha$.
   We will show that this produces precisely the terms required by Einstein's 
   Equation. The logarithm generates something new which was not present in the
   classical solution - a term of order $G\alpha\hbar / mr^3$. Here we have
   reinserted
   powers of $\hbar$ to emphasize that this is a quantum correction.

   It is interesting that a Feynman diagram calculation can generate the 
   {\em classical} correction in the metric. However, we will demonstrate
 that this is simply the 
   long-range electromagnetic field which surrounds a charged 
   particle. The logarithm comes from the same 
 class of Feynman diagrams - it encompases 
   the quantum fluctuations of the long-range field. The fact that it is the
   {\it long-range}
   component that determines the non-analytic behavior indicates that the 
   internal 
 structure of the particle is not relevant. Short range internal structure, for
   example using a proton instead of an electron, can be represented by a Taylor
   series
   in the form factor - and hence only involve analytic terms.

   The non-analytic terms are unambiguous finite effects in QED. How can we be
   sure that other quantum effects are not larger than these? Again the logic
   of effective field theory allows us to answer this. Quantum effects of
   massive degrees of freedom are always short ranged at low energy. Hence massive
   fields yield only analytic terms. The only other relevant massless degrees of
   freedom are gravitons\footnote{ Should one or more of the neutrinos be strictly 
   massless, weak interaction effects could also produce long range modifications
    to the metric. However, dimensional analysis shows that these are
   smaller
   than photonic effects. } The long distance classical and 
   quantum gravity effects are also calculable from the non-analytic components using
   effective field theory techniques\cite{don}. Because the gravitational
   interaction
   has a dimensionful coupling, Newton's constant $G$, the power counting is 
   different with the corespondence\cite{don}
   \begin{eqnarray}
    {G \alpha \over r^2}  &\to& {G^2 m^2 \over r^2}  \\
   {G \alpha \hbar \over m r^3}  &\to& {G^2 m \hbar \over r^3}
   \end{eqnarray}
   Therefore, as long as $Gm^2 < \alpha$, the QED effects will be dominant
   over quantum gravity effects. Note that describing grantum gravity modifications
   by a change in the metric is not straightforward, since the long-range
   propagation
   of gravity also has a quantum modification. This will be addressed in a future 
   paper\cite{dhb}. 

   Having identified the non-analytic terms which yield the long-range
   modifications
   of the metric, we now turn to the extraction of these effects.

   \section{Extracting the classical and quantum corrections}

     Defining the
   metric tensor via
   \begin{equation}
   g_{\mu\nu}=\eta_{\mu\nu}+h_{\mu\nu}
   \end{equation}
   where $\eta_{\mu\nu}=(1,-1,-1,-1)_{\rm diag}$ is the usual Minkowski
   metric, the interaction Hamiltonian has the form\cite{wein}
   \begin{equation}
   H=\int d^3x{1\over 2}T^{\mu\nu}(x)h_{\mu\nu}(x)\label{eq:grav}
   \end{equation}
   where $T^{\mu\nu}(x)$ is the energy-momentum tensor.  The analog of the
   Maxwell equation is the (linearized) Einstein equation, which has the
   form, in harmonic gauge---$\partial^\mu h_{\mu\nu}={1\over
   2}\partial_\nu h^\mu_\mu$---
   \begin{equation}
   \Box h_{\mu\nu}=-16\pi G(T_{\mu\nu}-{1\over 2}\eta_{\mu\nu}
   T)\label{eq:ein}
   \end{equation}
where $T=\eta^{\mu\nu}T_{\mu\nu}$ is the trace. 
The metric for a nearly static source is then recovered via the Green function
in either coordinate or momentum space
\begin{eqnarray}
h_{\mu\nu}(x) &=& -16\pi G \int d^3y D(x-y) 
(T_{\mu\nu}(y)-{1\over 2}\eta_{\mu\nu}
   T(y))  \\
&=& -16\pi G \int {d^3q	\over (2\pi)^3} e^{i\vec{q}\cdot\vec{r}}{1\over \vec{q}^2}(T_{\mu\nu}(q)-{1\over 2}\eta_{\mu\nu}
   T(q)) 
\end{eqnarray}

\subsection{Spinless particles}

   For a quantum mechanical system, $T_{\mu\nu}$ is represented by the
   transition density
   $$<p_2|T_{\mu\nu}(x)|p_1>$$
   and the conservation condition $\partial^\mu T_{\mu\nu}=0$ together
   with the requirement that $T_{\mu\nu}$ transform as a second rank
   tensor demands the general (scalar field) form
   \begin{equation}
   <p_2|T_{\mu\nu}(x)|p_1>={e^{i(p_2-p_1)\cdot x}\over
   \sqrt{4E_2E_1}}\left[2P_\mu P_\nu F_1(q^2)+(q_\mu q_\nu-g_{\mu\nu}
   q^2)F_2(q^2)\right]
   \end{equation}
   where $q_\mu = (p_2-p_1)_\mu$ is the momentum transfer and $q^2= q_0^2-\vec{q}^2$.
As can be seen from the condition
\begin{equation}
<p_2|\hat{P}_\mu|p_1>=<p_2|\int d^3xT_{\mu 0}(x)|p_1>=P_\mu<p_2|p_1>
\end{equation}
conservation of energy-momentum 
   requires
   $F_1(q^2=0)=1$
   but there exists no such constraint on $F_2(q^2)$. In QED at lowest order in $\alpha$ we have $F_1=1$ and $F_2= -1/2$.

\begin{figure}
\begin{center}
\epsfig{file=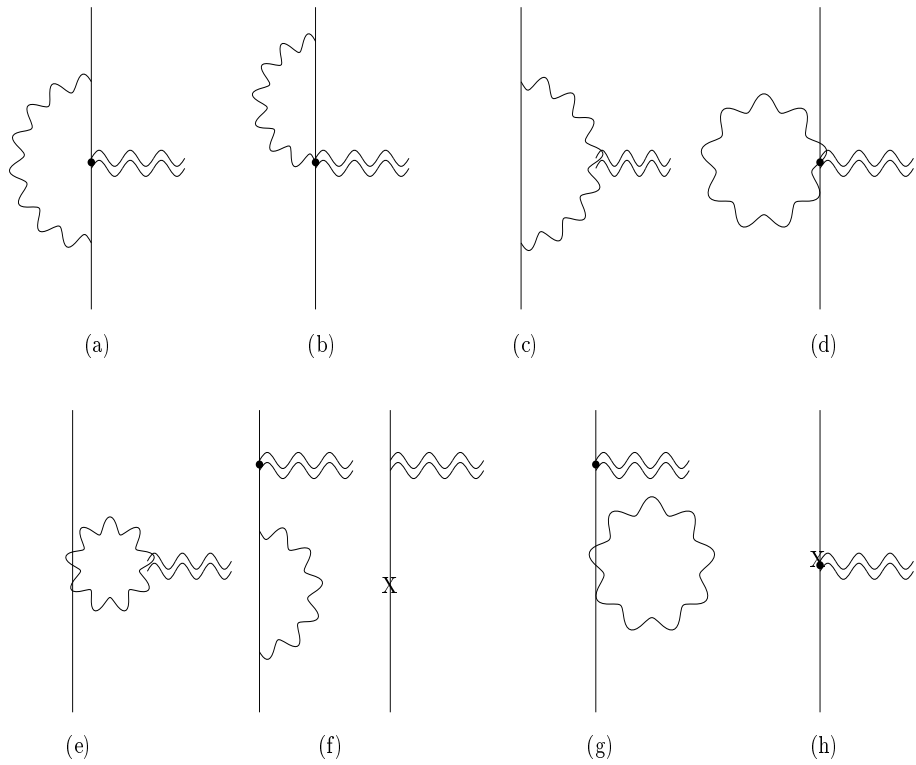,height=10cm,width=12cm}
\caption{Feynman diagrams for spin 0 radiative corrections to $T_{\mu\nu}$.}
\end{center}
\end{figure}

In order to calculate these form factors in QED, we calculate the diagrams of 
Fig. 1. Although we display the result of the full calculation below, the only diagrams relevant for the non-analytic terms are Figs. 1c and 1e,  where the graviton couples to the photon. This are the only diagrams with long range progagation from a massless particle. The form factors near $q^2=0$
are found to be
\begin{eqnarray}
F_1(q^2)&=&1+{\alpha\over 4\pi}{q^2\over m^2}\left(-{8\over 3}+{3\over
4}{m\pi^2\over \sqrt{-q^2}}+2\log{-q^2\over m^2}-{4\over
3}\log{\lambda\over m}\right)+\ldots\nonumber\\
F_2(q^2)&=&-{1\over 2}+{\alpha\over 4\pi}
\left(-\Omega-{26\over 9}+{m\pi^2\over 2\sqrt{-q^2}}+{4\over 3}\log{-q^2\over
m^2}\right)+\ldots\label{eq:ff1}
\end{eqnarray}
where 
$$\Omega={2\over \epsilon}-\gamma-\log{m^2\over 4\pi\mu^2}$$
is an ultraviolet divergence, which can be absorbed into the 
coeffcient of a term 
\begin{equation}
{\cal L}=KRF_{\mu\nu}F^{\mu\nu}{\rm tr}QUQU^\dagger\,.
\end{equation}
in the effective Lagrangian.  
There are a number of features here which are worthy of note.  One is
that the $q^2=0$ value of the leading
form factor $F_1(q^2)$ is unchanged from its lowest order size, as
required by energy-momentum conservation, while the form
factor $F_2(q^2)$ {\it is} modified at $q^2=0$.
However, the most important new effect is the
appearance of nonanalytic terms $\sim\sqrt{-q^2},q^2\log -q^2$, etc. in
the form factors. The square roots are associated with Fig. 1c only, in which the massive scalar can exist close to its mass shell, while the logarithms 
come from both Fig. 1c and 1e. 

We can explicitely demonstrate that the non-analytic terms are associated
with
long-range components of the energy momentum tensor by transformation to the coordinate space representation
\begin{equation}
T_{\mu\nu}(\vec{r}) = \int {d^3q\over (2\pi)^3}e^{i\vec{q}\cdot\vec{r}} T_{\mu\nu}(\vec{q})
\end{equation}
We work in the Breit frame where $q_0=0$ and $\vec{p}_2=-\vec{p}_1 =\vec{q}/2$.
The general relationships are
\begin{eqnarray}
    T_{00}(\vec{r})&=&\int{d^3q\over
   (2\pi)^3}e^{i\vec{q}\cdot\vec{r}}\left(mF_1(-\vec{q}^2)+
   {\vec{q}^2\over 2m}F_2(-\vec{q}^2)\right)\nonumber\\
    T_{0i}(\vec{r})&=&0\nonumber\\
   T_{ij}(\vec{r})&=&{1\over 2m}\int{d^3q\over
   (2\pi)^3}e^{i\vec{q}\cdot\vec{r}}(q_iq_j-\delta_{ij}\vec{q}^2)F_2(-\vec{q}^2)
\end{eqnarray}
This calculation involves a set of integrals which are listed in the 
Appendix.
Performing the Fourier transform for the one loop form factors
we find
\begin{eqnarray}
    T_{00}(\vec{r}) &=&\int{d^3q\over
   (2\pi)^3}e^{i\vec{q}\cdot\vec{r}}
\left(m-{\alpha\pi |\vec{q}| \over 8}-{\alpha \vec{q}^2\over 3\pi m}\log
   \vec{q}^2 \right)+\ldots\nonumber \\
&=&m\delta^3(\vec{r})+{\alpha\over 8\pi r^4} -{\alpha\hbar\over \pi^2mr^5}+\ldots\nonumber\\
    T_{0i}(\vec{r})&=&0\nonumber\\
   T_{ij}(\vec{r}) &=&\int{d^3q\over
   (2\pi)^3}e^{i\vec{q}\cdot\vec{r}}(q_iq_j-\delta_{ij}\vec{q}^2)
\left( {\alpha\pi\over 16 |\vec{q}|} +{\alpha\over 6\pi
   m}\log \vec{q}^2 \right)+\ldots\nonumber\\
   &=&-{\alpha\over 4\pi r^4}\left({r_ir_j\over r^2}-{1\over 2}\delta_{ij}\right)-
{\alpha \hbar \over 3\pi^2mr^5}\delta_{ij}+\ldots\label{eq:tg}
   \end{eqnarray}
Short range modifications of the matter distribution may smear out the delta function but will not change 
the long range fields.  The ``physics'' origin of these quantum corrections can be understood in terms of
the position uncertainty associated with quantum meachanics which implies the replacement of the 
distance $r$ in the classical expression by the value $\sim r+{\hbar\over m}$.  Since for macroscopic
distances $\hbar/m<<r$ expansion of he classical result in powers of $1/r$ yields to the form of the
quantum modifications found in our one loop calculation. 

Even though these forms for the energy momentum tensor were calculated from Feynman diagrams, we can verify that what we have called 
the classical component does in fact represent the classical energy momentum contained in the electric field around a charged particle.
 Since
for electromagnetism\cite{jack}
\begin{equation} 
T_{\mu\nu}^{EM}=-F_{\mu\lambda}F_\nu{}^\lambda+{1\over 4}\eta_{\mu\nu}F_{\lambda\delta}
F^{\lambda\delta}\label{eq:emt}
\end{equation}
we expect the 
energy momentum
tensor in the region around a charged mass to be
\begin{eqnarray}
 T_{00}^{EM}(\vec{r})&=&{1\over 2}E^2={\alpha\over 8\pi r^4}\nonumber\\
 T_{0i}^{EM}(\vec{r})&=&0\nonumber\\
 T_{ij}^{EM}(\vec{r})&=&-E_iE_j+{1\over 2}\delta_{ij}E^2=-{\alpha\over 4\pi
r^4}\left({r_ir_j\over r^2}-{1\over 2}\delta_{ij}\right)
\end{eqnarray}
This demonstrates the equivalance of the square-root non-analytic terms in the 
form factor with the classical field surrounding the particle. The logarithmic terms clearly yield the energy momentum of quantum fluctuations in this field.

From the form factors we may also directly reproduce the metric. Here the metric is determined from the two form factors by
\begin{eqnarray}
    h_{00}(\vec{r})&=&-16\pi G\int{d^3q\over
   (2\pi)^3}e^{i\vec{q}\cdot\vec{r}}{1\over\vec{q}^2}\left({m\over 2}F_1(-\vec{q}^2) -
{\vec{q}^2 \over 4m}F_2(-\vec{q}^2) \right) \nonumber\\
   h_{0i}(\vec{r})&=&0\nonumber\\
   h_{ij}(\vec{r})&=&-16\pi G\int{d^3q\over
   (2\pi)^3}e^{i\vec{q}\cdot\vec{r}}{1\over \vec{q}^2}
\left( {m\over 2}F_1(-\vec{q}^2)\delta_{ij} +{1\over 2m}(q_iq_j + {1\over 2} \delta_{ij}
\vec{q}^2) F_2(-\vec{q}^2)\right)\nonumber\\
\quad
   \end{eqnarray}
The leading constant term in the form factor obviously reproduces the usual Newtonian term in the metric.
The analytic terms reproduce a Dirac delta function, or derivitives of delta functions. These have no long range components and we do not display the results. Again the required integrals are collected in the Appendix.
The results using the one loop form factors are
\begin{eqnarray}
    h_{00}(\vec{r}) &=&-16\pi G\int{d^3q\over
   (2\pi)^3}e^{i\vec{q}\cdot\vec{r}}{1\over
\vec{q}^2} \left({m\over 2} -{\alpha\pi |\vec{q}|\over 8 } - {\alpha \vec{q}^2\over 3\pi m}\log
   \vec{q}^2 \right)+\ldots \nonumber\\
&=& -{2Gm\over r}+{G\alpha\over r^2}-{8G\alpha\hbar\over 3\pi mr^3}+\ldots\nonumber\\
   h_{0i}(\vec{r})&=&0\nonumber\\
   h_{ij}(\vec{r})&=&-16\pi G\int{d^3q\over
   (2\pi)^3}e^{i\vec{q}\cdot\vec{r}}{1\over \vec{q}^2}
\left(\delta_{ij}{m\over 2} +{\alpha\pi\over 16|\vec{q}|}
\left(q_iq_j-\delta_{ij}\vec{q}^2\right)\right.\nonumber\\
 &+&\left.{\alpha \over 6\pi m} ({q_iq_j}-\delta_{ij}\vec{q}^2)
\log \vec{q}^2\right)+\ldots \nonumber\\
   &=& -\delta_{ij}{2Gm\over r}+G\alpha{r_ir_j\over
   r^4}+{4G\alpha\hbar\over 3\pi mr^3}({r_ir_j\over r^2}-\delta_{ij})+\ldots\nonumber\\
   \quad
   \end{eqnarray}
These are precisely the appropriate forms for the Reissner-Nordstr\"{o}m 
metric---Eq. \ref{eq:rn}---along with the associated quantum corrections. 

\subsection{Spin 1/2}

 Let us now turn our attention to the case of a particle with spin, in particular spin one-half.
 The general form for the spin 1/2 matrix element of the
   energy-momentum tensor can be written as\cite{pag}
   \begin{eqnarray}
   <p_2|T_{\mu\nu}|p_1>&=&\bar{u}(p_2)\left[ F_1(q^2)P_\mu P_\nu{1\over
   m}\right. \nonumber\\
   &-&F_2(q^2)({i\over 4m}\sigma_{\mu\lambda}q^\lambda
   P_\nu+{i\over 4m}\sigma_{\nu\lambda} q^\lambda P_\mu)\nonumber\\
   &+& \left. F_3(q^2)(q_\mu q_\nu-g_{\mu\nu}q^2){1\over m}\right] u(p_1)
   \end{eqnarray}
The normalization condition $F_1(q^2=0)=1$ corresponds to energy-momentum 
conservation as found before, while the second normalization condition $F_2(q^2=0)=1$ is required by the constraint of angular momentum conservation.  This
can be seen by defining 
\begin{eqnarray}
\hat{M}_{12}&=&\int d^3x(T_{01}x_2-T_{02}x_1)\nonumber\\
&&\stackrel{q\rightarrow 0}{\longrightarrow}-i\nabla_{q_2}\int d^3x
e^{i\vec{q}\cdot\vec{r}}T_{01}(\vec{r})
+i\nabla_{q_1}\int d^3xe^{i\vec{q}\cdot\vec{r}}T_{02}(\vec{r})
\end{eqnarray} 
Then we find
\begin{equation}
\lim_{q\rightarrow 0}<p_2|\hat{M}_{12}|p_1>={1\over 2}={1\over 2}
\bar{u}_\uparrow(p)\sigma_3u_\uparrow(p)F_2(q^2)
\end{equation}
{\it i.e.}, $F_2(q^2=0)=1$, as found explicitly in our calculation.

\begin{figure}
\begin{center}
\epsfig{file=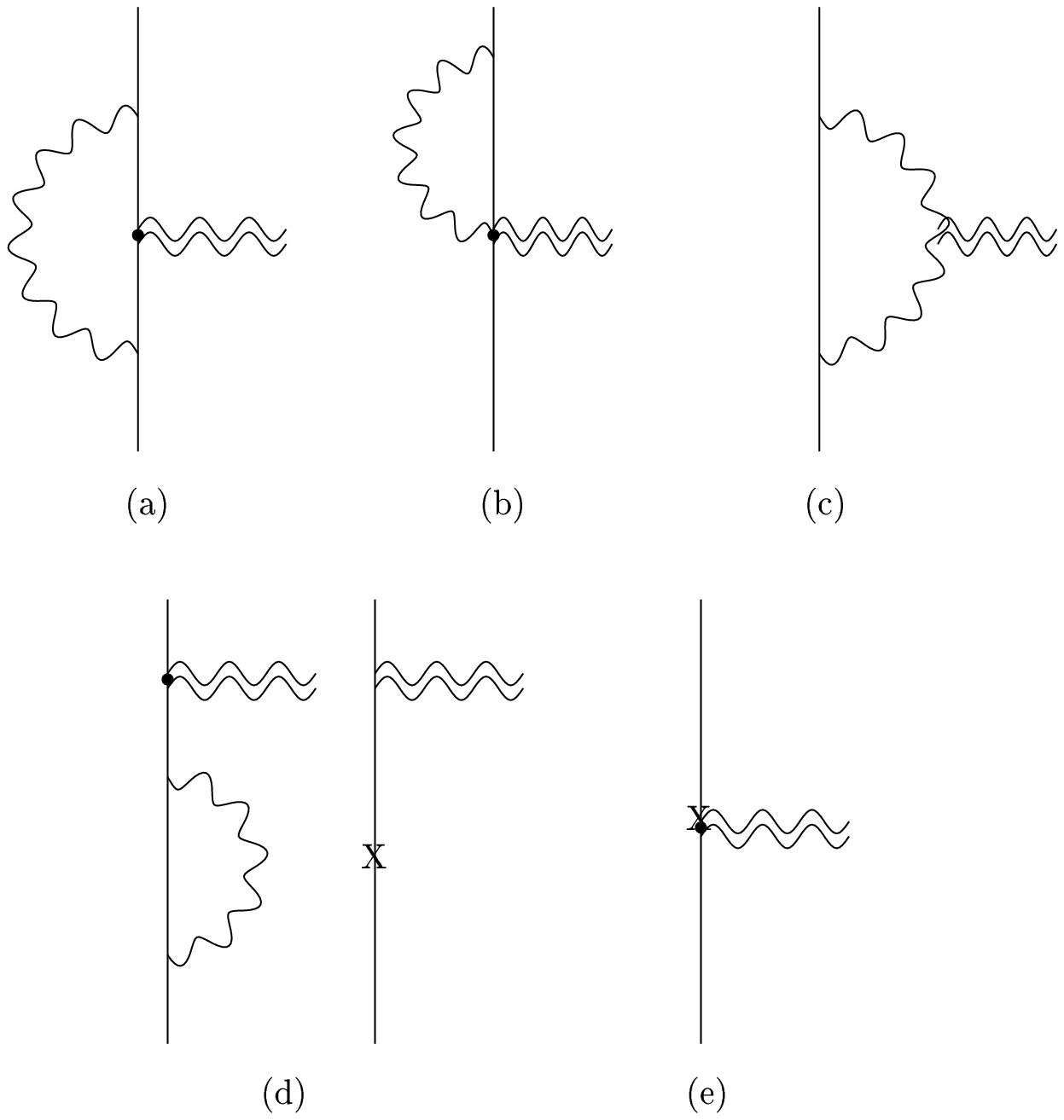,height=8cm,width=10cm}
\caption{Feynman diagrams for spin 1/2 radiative corrections to $T_{\mu\nu}$.}
\end{center}
\end{figure}

  The Feynman diagrams for fermions are shown in Fig 2. In this case only
one diagram, Fig. 2c, will be relevant for the non-analytic terms.  We
find
   \begin{eqnarray}
   F_1(q^2)&=&1+{\alpha\over 4\pi}{q^2\over m^2}\left( -{39\over
   18}+{3\pi^2m\over 4\sqrt{-q^2}}+2\log{-q^2\over m^2}-{4\over
   3}\log{\lambda\over m}\right) +\ldots \nonumber\\
   F_2(q^2)&=&1+{\alpha\over 4\pi}{q^2\over m^2}\left( -{47\over 18}+{1\over
   2}{\pi^2m\over \sqrt{-q^2}}+{2\over 3}\log{-q^2\over m^2}-{4\over
   3}\log{\lambda\over m}\right) +\ldots\nonumber\\
   F_3(q^2)&=&{\alpha\over 4\pi}\left( -{11\over 18}+{\pi^2m\over
   4\sqrt{-q^2}}+{2\over 3}\log{-q^2\over m^2}\right) +\ldots\label{eq:ff2}
   \end{eqnarray}

We convert this into an energy-momentum tensor.  Writing $\vec{S}=
\vec{\sigma}/2$ for the spin, the general relation to
the fermion form factors is
\begin{eqnarray}
    T_{00}(\vec{r})&=&\int{d^3q\over
   (2\pi)^3}e^{i\vec{q}\cdot\vec{r}}\left(mF_1(-\vec{q}^2)+
   {\vec{q}^2\over m}F_3(-\vec{q}^2)\right)\nonumber\\
    T_{0i}(\vec{r})&=&i\int{d^3q\over
   (2\pi)^3}e^{i\vec{q}\cdot\vec{r}}{1\over
   2}(\vec{S}\times\vec{q})_iF_2(-\vec{q}^2)\nonumber\\
   T_{ij}(\vec{r})&=&{1\over m}\int{d^3q\over (2\pi)^3}e^{i\vec{q}\cdot\vec{r}}
(q_iq_j-\delta_{ij}\vec{q}^2)F_3(-\vec{q}^2)\label{eq:tc}
   \end{eqnarray}
   The calculation may be performed using the integrals listed in the
   Appendix 
and we find
   \begin{eqnarray}
   T_{00}(\vec{r})&=&\int{d^3q\over
   (2\pi)^3}e^{i\vec{q}\cdot\vec{r}}\left( m -{\alpha\pi\over
   8}|\vec{q}|  
-{\alpha \over 3\pi m}
\vec{q}^2\log \vec{q}^2\right)+\ldots \nonumber \\
&=&m\delta^3(\vec{r}) +{\alpha\over 8\pi r^4}-{\alpha\hbar\over 
\pi^2m^2r^5}+\ldots\nonumber\\
    T_{0i}(\vec{r})&=&{i\over 2}\int{d^3q\over (2\pi)^3}
e^{i\vec{q}\cdot\vec{r}}(\vec{S}\times\vec{q})_i 
\left(1-{\alpha\pi\over
   8m}|\vec{q}| - {\alpha\over 6\pi m^2}\vec{q}^2\log \vec{q}^2
   \right)
+\ldots\nonumber \\
&=& {1\over 2}(\vec{S}\times\vec{\nabla})_i \delta^3(\vec{r})
+\left(-{\alpha\over 4\pi mr^6} +{5\alpha\hbar\over
4\pi^2m^2r^7} \right)(\vec{S}\times\vec{r})_i+\ldots \nonumber\\
   T_{ij}(\vec{r})&=&\int{d^3q\over
   (2\pi)^3}e^{i\vec{q}\cdot\vec{r}}\left(
{\alpha\pi\over
   16|\vec{q}|}+{\alpha\over 6\pi
m}\log \vec{q}^2 \right)\left(q_iq_j-\delta_{ij}\vec{q}^2\right)+\ldots
\nonumber \\
&=&-{\alpha\over 4\pi r^4}\left({r_ir_j\over r^2}-{1\over
   2}\delta_{ij}\right) -{\alpha\hbar\over
3\pi^2mr^5}\delta_{ij}+\ldots
   \quad\label{eq:tb}
\end{eqnarray}
As expected, the classical fields are the same as for the scalar case. 
The fact that the spin-independent quantum terms are also the same seems to be 
reasonable if they represent the quantum fluctuations of the electromagnetic fields. 
This result is, however, non-trivial in the Feynman diagram
calculation as 
different diagrams are involved. 
The form of the spin-dependent corrections can also be understood via 
simple classical arguments.
Since the time-space component of the energy-momentum tensor is given by Eq. 
\ref{eq:emt} as
\begin{equation}
T_{0i}=-(\vec{E}\times\vec{B})_i
\end{equation}
if we combine the electric field from a point charge as before 
with the magnetic field which arises
from a spinning particle with gyromagnetic ratio $g$
\begin{equation}
\vec{B}={eg\over 2m}{3\hat{r}\vec{S}\cdot\hat{r}-\vec{S}\over 4\pi r^3}
\end{equation}
the associated classical value of $T_{0i}$ is found to be
\begin{equation}
T_{0i}=-{\alpha g\over 8\pi mr^6}(\vec{S}\times\vec{r})_i
\end{equation}
Obviously agreement with Eq. \ref{eq:tb} is found if the Dirac 
value $g=2$ is used.

Similarly we can obtain the metric components due to this energy-momentum.
The relation of the metric to the fermion form factors is
 \begin{eqnarray}
    h_{00}(\vec{r})&=&-16\pi G\int{d^3q\over
   (2\pi)^3}e^{i\vec{q}\cdot\vec{r}}{1\over \vec{q}^2}\left( {m\over 2}F_1(-\vec{q}^2) -
{\vec{q}^2 \over 2m}F_3(-\vec{q}^2) \right)\nonumber\\
   h_{0i}(\vec{r})&=& -16\pi Gi\int{d^3q\over
   (2\pi)^3}e^{i\vec{q}\cdot\vec{r}}{1\over \vec{q}^2}{F_2(-\vec{q}^2) \over
   2}(\vec{S}\times\vec{q})_i \nonumber\\
   h_{ij}(\vec{r})&=&-16\pi G\int{d^3q\over
   (2\pi)^3}e^{i\vec{q}\cdot\vec{r}} {1\over \vec{q}^2}\left(  {m\over 2}F_1(-\vec{q}^2) +
{1\over m}(q_iq_j + {1\over 2} \delta_{ij}
\vec{q}^2) F_3(-\vec{q}^2) \right)\nonumber\\
\quad
   \end{eqnarray}
With the form factor calculated above this yields
\begin{eqnarray}
    h_{00}(\vec{r})&=&-16\pi G\int{d^3q\over
   (2\pi)^3}e^{i\vec{q}\cdot\vec{r}}{1\over \vec{q}^2}\left( {m\over 2} 
   -{\alpha\pi |\vec{q}|\over 8}     {\alpha \vec{q}^2\over 6\pi
   m}\log \vec{q}^2  
\right)+\ldots  \nonumber \\
   &=& -{2Gm \over r} +{G\alpha\over r^2}-{8G\alpha\hbar\over 3\pi
   mr^3}+
\ldots\nonumber\\
   h_{0i}(\vec{r})&=& -16\pi Gi\int{d^3q\over
   (2\pi)^3}e^{i\vec{q}\cdot\vec{r}}{1\over \vec{q}^2}\left({1\over 2}
   -{\alpha\pi |\vec{q}|\over
   16m} -{\alpha \vec{q}^2 \over 12\pi m^2}\log \vec{q}^2\right)
(\vec{S}\times\vec{q})_i +\ldots\nonumber \\
&=& \left( {2G\over r^3} -{G\alpha\over
   mr^4} + {2G\alpha\hbar\over \pi
   m^2r^5}\right)(\vec{S}\times\vec{r})_i
+\ldots\nonumber\\
   h_{ij}(\vec{r})&=&-16\pi G\int{d^3q\over
   (2\pi)^3}e^{i\vec{q}\cdot\vec{r}}{1\over \vec{q}^2}\left( {m\over 2} \delta_{ij} +
{\alpha\pi\over 16|\vec{q}|}
   (q_iq_j-\delta_{ij}\vec{q}^2)\right.\nonumber\\
&+&\left. {\alpha\over 6\pi m} (
{q_iq_j}-\delta_{ij}\vec{q}^2)\log \vec{q}^2\right)+\ldots\nonumber \\
&=& -\delta_{ij}{2Gm\over r}+{{G\alpha r_ir_j\over r^4}}   
+{4G\alpha\hbar\over 3\pi mr^3}\left({r_ir_j\over r^2}-
\delta_{ij}\right)+\ldots\label{eq:rc1}
   \end{eqnarray}
We observe that the diagonal components are identical to those found
   for the spinless case, as expected, and that there exists a
   nonvanishing non-diagonal term associated with the spin. 
 We have thus reproduced the Kerr-Newman metric in harmonic 
gauge---Eq. \ref{eq:kn}---together with the associated quantum corrections.

   \section{Conclusions}

   Above we have examined the radiative corrections to the lowest order
   gravitational coupling of a massive charged particle.  
 We have seen that the form of the $q^2$ dependence to
   these form factors has an important difference from the structure
   of most other form factors.  In addition to the usual analytic
   terms such as $q^2/m^2, etc.$, the gravitational form factors of 
charged particles also
   include nonanalytic components such as $\sqrt{-q^2},q^2\log -q^2,etc.$
   which are associated with the feature that the graviton can couple to
   a massless field---in this case the virtual photon.  By transforming
   to co-ordinate space, we demonstrated that these new forms determine
   long range corrections to the energy-momentum tensor and, via the
   Einstein equation, to the gravitational field, whose form and strength
   are fully constrained and determined by the feature that nature is
   describable in terms of a quantum field theory.  Specifically, leading
   corrections to the Schwarzschild solution of ${\cal O}(G\alpha/r^2)$ are found
   to
   correspond to well known classical solutions, and quantum corrections
   of order ${\cal O}(G\alpha\hbar/mr^3)$ are determined.  These terms
   appear 
too small to be measured. However, their
   presence is an interesting application of effective 
   field theoretic methods in the case of the gravitational interaction.

   Going further, we know that the photon is not the only massless field
   to which the graviton couples---self interaction associated with
   the nonlinear nature of
   gravity guarantees that the graviton also couples to itself.  Thus
   there exists an extension of this work to
   graviton loop diagrams and to higher order components of the
   gravitational self interaction.  Some of these have already been
   explored in earlier work by one of us\cite{don},
 but a more general discussion
   will be the subject of a future communication\cite{dhb}.

   \section{Appendix}

Here we collect the integrals that are needed in the Fourier transformation
of the non-analytic terms. For the calculation of the long range part
of the 
energy momentum tensor we use
\begin{eqnarray}
\int{d^3q\over (2\pi)^3}e^{i\vec{q}\cdot\vec{r}}|\vec{q}|=-{1\over
\pi^2r^4}&,&\quad \int{d^3q\over
(2\pi)^3}e^{i\vec{q}\cdot\vec{r}}q_j|\vec{q}|={-4ir_j\over \pi^2r^6}\nonumber\\
\int{d^3q\over
(2\pi)^3}e^{i\vec{q}\cdot\vec{r}}{q_iq_j\over |\vec{q}|}&=&{1\over
\pi^2r^4}\left(\delta_{ij}-4{r_ir_j\over r^2}\right)\label{eq:r1}
\end{eqnarray}
and
\begin{eqnarray}
   \int{d^3q\over (2\pi)^3}e^{i\vec{q}\cdot\vec{r}}\vec{q}^2\log 
\vec{q}^2={3\over
   \pi r^5}&,& \quad \int{d^3q\over
   (2\pi)^3}e^{i\vec{q}\cdot\vec{r}}{q_j\vec{q}^2\log \vec{q}^2}=
{i15r_j\over \pi
   r^7}\nonumber\\
   \int{d^3q\over
   (2\pi)^3}e^{i\vec{q}\cdot\vec{r}}q_iq_j\log \vec{q}^2&=&{1\over
   \pi r^5}\delta_{ij}\label{eq:r2}
   \end{eqnarray}
In calculating the metric we need
\begin{eqnarray}
   \int{d^3q\over
   (2\pi)^3}e^{i\vec{q}\cdot\vec{r}}{1\over |\vec{q}|}
={1\over 2\pi^2r^2}&,&\quad
   \int{d^3q\over
   (2\pi)^3}e^{i\vec{q}\cdot\vec{r}}{q_j\over |\vec{q}|}
={ir_j\over \pi^2r^4}\nonumber\\
   \int{d^3q\over
   (2\pi)^3}e^{i\vec{q}\cdot\vec{r}}{q_iq_j\over |\vec{q}|^3}&=&{1\over
   2\pi^2r^2}\left(\delta_{ij}-2{r_ir_j\over r^2}\right)\label{eq:r3}
   \end{eqnarray}
for the square-roots and
\begin{eqnarray}
   \int{d^3q\over
   (2\pi)^3}e^{i\vec{q}\cdot\vec{r}}\log \vec{q}^2=-{1\over 2\pi r^3}&,&\quad
   \int{d^3q\over
   (2\pi)^3}e^{i\vec{q}\cdot\vec{r}}q_j\log \vec{q}^2
={-i3r_j\over 2\pi r^5}\nonumber\\
   \int{d^3q\over
   (2\pi)^3}e^{i\vec{q}\cdot\vec{r}}\log \vec{q}^2\left({q_iq_j\over
   \vec{q}^2}\right)&=&-{r_ir_j\over 2\pi r^5}\label{eq:r4}
   \end{eqnarray}
for the logarithms. 
   \begin{center}
   {\bf Acknowledgement}
   \end{center}

   This work was supported in part by the National Science Foundation
   under award PHY-98-01875.  BRH would like to acknowledge the warm
hospitality of Forschungszentrum J\"{u}lich and the support of the
Alexander von Humboldt Foundation.


\begin{thebibliography}{99}

\bibitem{dgh} See, {\it e.g.}, J.F. Donoghue, E. Golowich, and
   B.R. Holstein, {\bf Dynamics of the Standard Model}, Cambridge
   Univ. Press, New York (1992).
   \bibitem{eft} See, {\it e.g.}, B.R. Holstein, in {\bf Hadron Physics
   2000}, ed. F.S. Navarra, M.R. Robilotta, and G. Krein, World
   Scientific, Singapore (2001).
   \bibitem{gl} J. Gasser and H. Leutwyler, Ann. Phys. (NY) {\bf 158},
   142 (1984); Nucl. Phys. {\bf B250}, 465 (1985).
 \bibitem{don} J.F. Donoghue, Phys. Rev. {\bf D50}, 3874 (1994).
 \bibitem{rnm} H. Reissner, Ann. Phys. {\bf 50}, 106 (1916);
   G. Nordstr\"{o}m, Proc. Kon. Ned. Akad. Wet. {\bf 20}, 1238 (1918).
  \bibitem{ker} R. Kerr, Phys. Rev. Lett. {\bf 11}, 237 (1963).
\bibitem{knm} E.T. Newman, E. Couch, K. Chinnapared, A. Exton,
   A. Prakash, and R. Torrence, J. Math. Phys. {\bf 6}, 918 (1965).
   \bibitem{ber} The calculations for spin 1/2 have previously performed
   by F. Berends, and R. Gastmans, Phys. Lett, {\bf B55}, 311 (1975);
   Ann. Phys. (NY) {\bf 98}, 225 (1976) and by K. Milton, Phys. Rev. {\bf D15},
   538 (1977).  In the case of spin 0 previous evaluations were
   done by B. Kubis and U.-G. Meissner, Nucl. Phys. {\bf A671}, 332
   (2000) and  Nucl. Phys. {\bf A692}, 647 (2001).  However, except for a
   brief discussion by Berends and Gastman the implications of the
   low $q^2$ region and its connection with coordinate space were not
   explored in these works.
   \bibitem{pagli} L.F. Li and H. Pagels, Phys. Rev. Lett. {\bf 26}, 1204 (1971); H. Pagels, Phys. Rept. {\bf C16}, 219 (1975). 
\bibitem{dhb} J.F. Donoghue, B.R. Holstein, and N.E.J. Bjerrum-Bohr, 
in preparation.
 \bibitem{wein} See, {\it e.g.}, S. Weinberg, {\bf Gravitation and
   Cosmology}, Wiley, New York (1972).
   \bibitem{jack} See, {\it e.g.}, J.D. Jackson, {\bf Classical
   Electrodynamics}, Wiley, New York (1975).
    \bibitem{pag} H. Pagels, Phys. Rev. {\bf 144}, 1250 (1966).

   \end{thebibliography}
   \end{document}